\documentstyle{l-aa}
\begin{document}

  \thesaurus{02.          
              (08.14.1;   
               02.04.1;   
               02.13.1;   
               02.18.6)   
            }
\title{
Electron conduction along quantizing magnetic fields 
in neutron star crusts}
\subtitle{
I. Theory
}
\author{A.Y.\,Potekhin}
\institute{A.F.\,Ioffe Physical-Technical Institute,
           194021, St-Petersburg, Russia\thanks{Permanent address} 
         \and 
           Nordita, Blegdamsvej 17, DK-2100 Copenhagen \O, Denmark}

\date{Received 7 April 1995 / Accepted 17 June 1995}

\maketitle

\begin{abstract}
Transport properties of degenerate relativistic electrons
along quantizing magnetic fields in neutron star crusts
are considered.
A kinetic equation is derived for the spin polarization
density matrix of electrons.
Its solution does not depend on
the choice of basic electron wave functions unlike previous
solutions of the traditional kinetic equation for the
distribution function. The density matrix formalism shows
that one can always reach high accuracy with
the traditional method
by a proper choice of the basic functions.
Electron Coulomb scattering on ions is considered in liquid matter, and
on high-temperature phonons or on charged impurities
in solid matter.
In the solid regime, the
Debye -- Waller reduction of phonon scattering
can strongly enhance the longitudinal thermal or electric conductivity.
An efficient numerical method is proposed for calculating
the transport properties of electron gas at any magnetic field
of practical interest.

\keywords{stars: neutron -- dense matter -- 
magnetic fields -- radiation mechanisms: thermal}
          
\end{abstract}
\section{Introduction}                                        
\label{sect1}
Accurate transport coefficients in neutron star
crusts are important for analysing the thermal
evolution of neutron stars and evolution of their magnetic fields.
In outer crusts of cooling magnetized neutron stars, 
the heat is mainly transported 
along the magnetic fields. There exist several 
competing heat transport mechanisms across the field, 
but the longitudinal currents are carried mostly by 
electrons through their scattering on phonons or charged 
impurities in the solid phase and on ions in the liquid phase. 
As a rule, the electrons in the crust are 
strongly degenerate and may be 
relativistic; the magnetic field can be easily quantizing. 
Transport properties of the crusts 
have been studied in a number of papers (e.g., Yakovlev \& Kaminker 1994, 
and references therein). 

In the present work, the most important 
problem of longitudinal electron transport in quantizing 
magnetic fields is studied with the use of 
the quantum density matrix formalism, 
instead of the traditional kinetic equation 
for the electron distribution function employed in previous studies. 
The main advantage of the present approach is that it is independent of 
the choice of the basis of electron states (basis states are not 
unique due to the electron spin degeneracy). 

We consider three main electron scattering mechanisms. 
The first one is the 
Coulomb scattering on ions in the liquid or gaseous phase. 
The second one is the scattering on high-temperature phonons 
in the solid phase. In the latter case, we take into account the 
Debye -- Waller factor whose importance 
has been emphasized and proved by Itoh et al. (1984b, 1993) 
for the non-magnetic case. We show that the effect of 
this factor is much stronger in quantizing magnetic fields. 
The third mechanism is the Coulomb scattering on charged impurities 
in the solid phase, important much below the melting temperature. 

The paper is composed as follows.
In Sect.~\ref{sect2} we describe the physical conditions
of interest, electron scattering potentials, transport
coefficients and their expressions in the non-magnetic case.
In
Sect.~\ref{sect3} we derive a linearized kinetic equation for the
density matrix and compare its numerical solutions with
the traditional solutions employed
in all previous works. Mathematical properties of the new
equation are discussed in Appendix~A. The effect of the Debye --
Waller factor in quantizing magnetic fields is studied in
Sect.~\ref{sect4}. The results are summarized in Sect.~\ref{sect5}. In
Appendix~B we present new expressions for some intermediate
integrals. These expressions ensure
efficient computation of
the transport properties for the case
when many Landau orbitals are
occupied. Previous results 
(Yakovlev 1984, Hernquist 1984, Schaaf 1988, Van Riper 1988)
were restricted to 30 Landau orbitals at most.

\section{Basic equations}                                 
\label{sect2}
\subsection{Physical conditions}       
Magnetic fields $B$ in magnetized
neutron star crusts are known to range between $10^{11}$
and $10^{14}$~G. For temperatures $T\sim (10^6-10^8)$~K
and densities $ \la 10^8$ g~cm$^{-3}$ of interest, the 
Landau 
quantization of electron motion in a magnetic field
can be very important (then the field is
called {\it quantizing\/}).
Consider not too low densities (typically, $ \ga 10^5$ g~cm$^{-3}$
at $B\sim 10^{12}$~G, see, e.g.,  Yakovlev 1984, Van Riper 1988)
when the electron gas is almost free and the ions are fully ionized.
An electron energy is then given by
\begin{equation} 
\epsilon_n(p_z) = c\,\left((m_{\rm e} c)^2+2\hbar\omega_B
m_{\rm e} n+p_z^2\right)^{1/2},
\label{2.0a}
\end{equation}
where $p_z$ is the electron momentum
along the field, $\omega_B=eB/(m_{\rm e}c)$ is the
electron cyclotron frequency, $(-e)$ is the electron charge,
$m_{\rm e}$ is the electron mass, $c$ is the speed of light, and 
$n=0,1,2,\ldots$~ is a Landau quantum number.

The electron number density can be expressed as
\begin{equation}
   n_{\rm e} = {m_{\rm e}\omega_B\over (2\pi\hbar)^2}
   \int_{-\infty}^\infty{\rm d}p_z\sum_{n,s} f_0(\epsilon_n(p_z)), 
\label{2.0}
\end{equation}
where $s$ is a spin quantum number
($s=-1$ for $n=0$, and $s=\pm1$ for $n=1,2,\ldots$, 
see Sect.~\ref{sect3}),
\begin{equation}
   f_0(\epsilon) = \left[\exp \left( {\epsilon - \mu
   \over k_{\rm B}T} \right)
   +1\right]^{-1}
\label{3.11}
\end{equation}
is the Fermi -- Dirac distribution function, and $\mu$ 
is the chemical potential (including $m_{\rm e}c^2$). 
A fitting formula for calculation of $n_{\rm e}$ is given in 
Appendix~C. 

The state of ions is mainly determined by the ion
coupling parameter
$\Gamma=(Ze)^2/(ak_{\rm B}T)$, where $Ze$ is an ion charge,
$k_{\rm B}$ is the Boltzmann constant, $a=(4\pi n_{\rm i}/3)^{-1/3}$ is 
the ion sphere radius, and $n_{\rm i}=n_{\rm e}/Z$ 
is the ion number density. 
If $\Gamma \ll 1$, the ions constitute an ideal gas.
For higher $\Gamma$, the gas gradually transforms into
a strongly coupled Coulomb liquid. The liquid solidifies
at the melting temperature $T=T_{\rm m}$ which corresponds 
to $\Gamma=172$ (Nagara et al. 1987). The zero-point quantum 
vibrations of the Coulomb crystal become comparable with 
the thermal vibrations
at $k_{\rm B}T \approx \hbar \omega_{\rm p}$, where 
$\omega_{\rm p} = \sqrt{4 \pi Z^2 e^2 n_{\rm i}/m_{\rm i}}$ 
is the ion plasma
frequency, and $m_{\rm i}$ is the ion mass. We will not consider 
superstrong magnetic fields (discussed, e.g., by Yakovlev
1984) which affect the properties of the ion plasma component.

A more detailed description of the physical conditions
is given by Yakovlev (1984) and Van Riper (1988).
\subsection{Scattering potentials}       
Consider three important cases when the electron scattering is 
almost elastic. The first case is the Coulomb scattering
on ions in the liquid or gaseous phase ($T > T_{\rm m}$).
The second case is the scattering on high-temperature phonons 
($\hbar\omega_{\rm p}/k_{\rm B} \la T < T_{\rm m}$). 
The third one is the Coulomb scattering 
on charged impurities in the lattice, which is 
important for $k_{\rm B}T \ll \hbar\omega_{\rm p}$. 
The impurities represent 
ions of charge $Z_{\rm imp} \neq Z$ immersed accidentally in 
lattice sites. 
Thus our results will cover a wide range of temperatures. 

The electron-ion 
scattering potential $V({\bf r})$ in the liquid or gaseous 
phase can approximately be taken as a screened 
Coulomb potential (Yakovlev 1984). 
Its Fourier image $U({\bf q})$ 
is given by 
\begin{equation}
   \left| U_{\rm ion}({\bf q})\right|^2 =
   \left[ 4\pi Ze^2 /(q^2+r_{\rm s}^{-2}) \right]^2,
\label{2.1}
\end{equation}
where $r_{\rm s}$ is an effective screening length, 
$r_{\rm s}^{-2}=r_{\rm i}^{-2}+r_{\rm e}^{-2}$. Here $r_{\rm i}$ 
and $r_{\rm e}$ 
are the screening lengths due to ions and electrons, respectively. 
In the most important liquid regime ($ 1 \la \Gamma < 172$), 
according to Yakovlev (1984), the ion screening length is 
\begin{equation}
   r_{\rm i} = a\sqrt{{\rm e}/6}\approx 0.67a 
\label{2.2}
\end{equation}
(Hernquist (1984) used a less accurate approximation for $r_{\rm i}$). 
The electron screening length is determined as 
\begin{equation}
r_{\rm e}=\left[4\pi e^2\;\partial n_{\rm e}/\partial\mu
\right]^{-1/2}. 
\label{2.2a}
\end{equation}
When the temperature is low enough 
($k_{\rm B}T \ll \hbar\omega_B^\ast$ with 
$\omega_B^\ast = \omega_B m_{\rm e}c^2/\mu$), 
we have 
\begin{eqnarray}
(a_{\rm m}/r_{\rm i})^2 &=& 
{6\over {\rm e}}\,\left(m_{\rm e}\hbar\omega_B\right)^{-1/3}\,
\left[{2\over 3\pi Z} \sum_{ns}|p_z|\right]^{2/3}, 
\label{2.2c}
\\
(a_{\rm m}/r_{\rm e})^2 &=& {2\alpha\over\pi}\,\sum_{ns} 
{\epsilon\over |p_z|c}, 
\label{2.2d}
\end{eqnarray}
where $a_{\rm m} = (\hbar c/eB)^{1/2}$ is the magnetic length, 
$\alpha=e^2/\hbar c$ is the fine-structure constant, 
and the energy and momentum variables are assumed to be 
taken on the Fermi surface: $\epsilon=\epsilon_n(p_z)=\mu$. 
At arbitrary temperature, 
the fitting formula of Appendix C can be used 
to calculate $n_{\rm e}$ and $\partial n_{\rm e}/\partial\mu$ 
for estimation of $r_{\rm i}$ and $r_{\rm e}$ 
according to Eqs.\,(\ref{2.2}) and (\ref{2.2a}). 

For scattering on high-temperature phonons in the solid phase,
one has
\begin{equation}
   \left|U_{\rm ph}({\bf q})\right|^2 =
   \left({4\pi Z e^2 \over q}\right)^2\,{r_T^2\over 3}\,
   \exp\left[-2W({\bf q})\right],
\label{2.3}
\end{equation}
where
\begin{equation}
   r_T^2 = { 3k_{\rm B}Tu_{-2}\over 4\pi n_{\rm i} Z^2 e^2} =
   {u_{-2}a^2\over\Gamma} 
\label{2.4}
\end{equation}
is the mean squared thermal displacement of ions, 
$u_{-2}$ is a numerical factor determined by the phonon 
spectrum ($u_{-2}=13$ for the bcc lattice), 
and ${\rm e}^{-2W}$ is the Debye -- Waller factor. 
The latter factor is usually negligible for scattering 
in terrestrial solids (e.g., Davydov 1976), 
but it is important 
in dense neutron star matter (Itoh et al. 1984b, 1993).
For the high-temperature solids 
($k_{\rm B}T \ga \hbar\omega_{\rm p}$) 
of interest, one has (e.g., Itoh et al. 1984a) 
\begin{equation}
   2W({\bf q}) \approx (r_T q)^2/3.
\label{2.4a}
\end{equation}
Note that Eq.\,(\ref{2.3}) represents the familiar high-temperature
asymptote of the one-phonon scattering potential
(Yakov\-lev \& Urpin 1980) multiplied by
the Debye -- Waller term (Itoh et al. 1984b) to include
multiphonon processes.

Finally, the Coulomb scattering on impurities corresponds to
(e.g., Yakovlev \& Urpin 1980)
\begin{equation}
   \left| U_{\rm imp}({\bf q})\right|^2 =
   \left[ 4\pi (Z_{\rm imp}-Z)e^2 /(q^2+r_{\rm s}^{-2}) \right]^2.
\label{2.1a}
\end{equation}
In this case the screening length $r_{\rm s}$ is most likely determined
by the electrons: $r_{\rm s}=r_{\rm e}$. 
The Coulomb scattering on impurities is very similar
to that on ions, and we do not consider them separately in detail. 
The results for impurities can be obtained 
from those for ions by replacing 
\begin{equation}
    Z \rightarrow Z_{\rm imp}-Z, ~~~ 
    n_{\rm i} \rightarrow n_{\rm imp},
\label{2.2b}
\end{equation}
where $n_{\rm imp}$ is the impurity number density.

Equations (\ref{2.1}) and (\ref{2.3}) can be conveniently written as 
\begin{eqnarray}
     \left|U_{\rm ion}({\bf q})\right|^2 &=&
     {2\pi\over n_{\rm i} l}\,
     \left[{2\hbar c/ a_{\rm m}\over q^2+r_{\rm s}^{-2}}\right]^2,
\label{2.5}
\\
    \left|U_{\rm ph}({\bf q})\right|^2 &=&
    {\pi\over n_{\rm i} l}\,\left({2\hbar c\over q}\right)^2\,
    {\rm e}^{-2W},
\label{2.6}
\end{eqnarray}
where $l$ is a scale length (Yakovlev 1984): 
\begin{equation}
    l_{\rm ion} = {m_{\rm e} c^2\hbar\omega_B \over 2\pi 
    n_{\rm i} Z^2 e^4},
    ~~~~~~
    l_{\rm ph} = {3\over 4\pi n_{\rm i}} 
    \left({\hbar c\over Ze^2 r_T}\right)^2. 
\label{2.8}
\end{equation}
For the scattering on impurities, $U_{\rm imp}$ and 
$l_{\rm imp}$ are obtained from $U_{\rm ion}$ and 
$l_{\rm ion}$ by using Eq.\,(\ref{2.2b}). 
\subsection{Transport coefficients}                
\label{sect2.3}
Let $j$ and $q$ be the densities of the electric and thermal
currents induced by sufficiently weak electric field
${\cal E}$ and gradients of temperature $T$ and electron chemical
potential $\mu$ directed along the magnetic field (along
the $z$-axis). The currents are determined by three transport
coefficients $\sigma$, $\beta$ and $\lambda$,
\begin{eqnarray}
   j &=& \sigma\left({\cal E}+{1\over e}\,
   {\partial\mu\over\partial z}\right)+
   \beta{\partial T\over\partial z},
\nonumber\\
   q &=& -\beta T\left({\cal E}+{1\over e}\,
   {\partial\mu\over\partial z}\right) -
   \lambda\,{\partial T\over\partial z},
\label{3.14}
\end{eqnarray}
where $\sigma$ is the longitudinal electric conductivity.

For practical use, Eq.\,(\ref{3.14})
can be rewritten as
\begin{equation}
   {\cal E} + {1 \over e} {\partial \mu \over \partial z}
    = {j \over \sigma} - {\beta\over\sigma} 
    {\partial T \over \partial z},
    ~~~~~ 
    q= - {\beta\over\sigma} T j - 
    \mbox{\ae} {\partial T \over \partial z},
\label{3.14a}
\end{equation}
where
$    \beta / \sigma$ and 
$    \mbox{\ae}= \lambda - T \beta^2 / \sigma$ 
are the longitudinal thermopower and thermal
conductivity, 
respectively. 

For nearly elastic electron scattering,
the kinetic coefficients $\sigma$, $\beta$, and $\lambda$
may be expressed as 
\begin{equation}
   \left(
   \begin{array}{c}
       \sigma \\ \beta \\ \lambda
   \end{array}
   \right) =
   \int_{m_{\rm e}c^2}^\infty\!
   {\left(
   \begin{array}{c}
       e^2 \\ e(\epsilon-\mu)/T \\ (\epsilon -\mu)^2/T
   \end{array}
   \right)
   {n_{\rm e}\tau(\epsilon) c^2 \over \epsilon}
   \left( -{\partial f_0\over\partial\epsilon} \right)
   {\rm d}\epsilon}, 
\label{3.15}
\end{equation}
where  $\tau(\epsilon)$ is the effective energy-dependent
relaxation time for the electrons. 
\subsection{Non-magnetic electron relaxation times}   
In the absence of the magnetic field (or for non-quantizing field), 
the inverse effective relaxation time          
(effective collision frequency) of an electron 
with energy $\epsilon$ can be presented as 
\begin{equation}
   \tau^{-1}(\epsilon) = n_{\rm i} v \sigma_{\rm tr}(\epsilon),
\label{4.2}
\end{equation}
where $v$ is the electron velocity
and $\sigma_{\rm tr}(\epsilon)$ is the transport cross section: 
\begin{equation}
   \sigma_{\rm tr}(\epsilon) = 
   \int {{\rm d}\Omega\over 4\pi}\int {\rm d}\Omega'
   \sigma({\bf p}\to {\bf p}')\,(1-\cos\Theta).
\label{4.3}
\end{equation}
Here 
${\bf p}$ and ${\bf p}'$ are electron momenta before and
after scattering, respectively, d$\Omega$ and d$\Omega'$ are
solid angle elements,
$\Theta$ is the scattering angle,
and $\sigma({\bf p}\to {\bf p}')$
is a differential scattering cross section.
In the Born approximation,
\begin{equation}
\sigma({\bf p}\to {\bf p}') = 
   {|U({\bf q})|^2\epsilon^2\over 4\pi^2\hbar^4c^4}\,
   \left(1-{v^2\over c^2}\,\sin^2{\Theta\over 2}\right).
   \label{4.3a}
\end{equation}
Consider the Coulomb scattering. Let us substitute 
the Fourier image (\ref{2.1}) into Eq.\,(\ref{4.3a}).
Integrating in Eq.\,(\ref{4.3}),
we arrive at the well known result (e.g., Yakovlev and Urpin 1980):
\begin{equation}
   \sigma_{\rm tr}(\epsilon) =
   4\pi\left({Ze^2\over pv}\right)^2
   \Lambda(\epsilon).  
\label{4.13}
\end{equation}
Here $\Lambda(\epsilon)$ is the Coulomb logarithm (Yakovlev 1980): 
\begin{eqnarray}
   \Lambda(\epsilon)
&=& {1\over 2}\,\left[\ln(1+w)-
(1+w^{-1})^{-1}\right] -
\nonumber\\ && 
{v^2\over 2c^2}\,\left[1-2w^{-1}
\ln\left(1+w\right)+
\left(1+w\right)^{-1}\right], 
\label{4.14a}
\end{eqnarray}
and $w\equiv (2pr_{\rm s}/\hbar)^2$. 
The terms of order $w^{-1}$ are often neglected 
in $\Lambda(\epsilon)$ (e.g., Yakovlev and Urpin 1980, 
Yakovlev 1984). 
However these terms are very significant in Coulomb liquids 
of carbon and lighter elements. 

Analogously, for the scattering on phonons, 
from Eqs.\,(\ref{2.3}) and (\ref{2.4a}) we obtain 
\begin{equation}
   \sigma_{\rm tr}(\epsilon) =
   {8\pi\over 3}\left({Ze^2\over\hbar v}\right)^2
   r_T^2\left(R_1(\epsilon)-R_2(\epsilon)\,{v^2\over 2c^2}\right),
\label{4.10}
\end{equation}
where 
\begin{equation}
   R_1 = {1\over w}\,\left(1-{\rm e}^{-w}\right), ~~~ 
   R_2 = {2\over w^2} \left(1-{\rm e}^{-w}\,(1+w)\right),
\label{4.10a}
\end{equation}
and $w \equiv (4/3)\,(pr_T/\hbar)^2$.
Generally, we have $R_1<1$ and $R_2<1$ due to
the Debye -- Waller factor. If $w \ll 1$,
the Debye -- Waller factor is insignificant,
$R_1=R_2=1$, and Eq.\,(\ref{4.10}) reproduces
the result of Urpin and Yakovlev
(1980).

Finally, one obtains
\begin{eqnarray}
   \tau_0^{\rm ion} &=& {p^2 v \over 
   4\pi\Lambda(\epsilon) Z^2 e^4 n_{\rm i}},
\label{4.10b}
\\
   \tau_0^{\rm ph} &=& {3\hbar^2 v \over 8\pi r_T^2 Z^2 e^4 n_{\rm i}}\,
   \left(R_1(\epsilon) - R_2(\epsilon)\,{v^2\over 2c^2}\right)^{-1},
\label{4.10c}
\end{eqnarray}
for the Coulomb and phonon scattering, respectively.

\section{Density matrix}                                      
\label{sect3}
\subsection{Kinetic equation}
\label{sect3.1}
Transport properties of a degenerate relativistic electron gas
in quantizing magnetic fields were studied by Yakovlev (1984),
Hernquist (1984), Van Riper (1988), and Schaaf (1988) on the basis
of the linearized kinetic equations for the electron distribution
function. However, as noted by Yakovlev (1984),
a selfconsistent description should be based on the quantum density
matrix formalism.

Let us consider a uniform magnetic field {\bf B} along the $z$-axis
and use the Landau gauge of the
vector potential: ${\bf A} = (-By, 0, 0)$. Quantum states of
a free electron in the magnetic field form a complete orthogonal
basis. The basic states can be labelled by the quantum numbers
$\kappa = (p_x, p_z, n, s)$, where 
$p_x$ determines the $y$-coordinate
of the guiding center, $y_B = p_x /(m_{\rm e} \omega_B)$.
An explicit solution of the Dirac equation
reads (e.g., Sokolov \& Ternov 1968) 
\begin{equation}
   \psi_\kappa({\bf r}) =
   {\exp[{\rm i}(p_xx+p_zz)/\hbar]\over
   (a_{\rm m} L_x L_z)^{1/2}} \,\chi_{ns}(p_z,y-y_B),
\label{3.1}
\end{equation}
where $L_x$ and $L_z$ are the normalization lengths, 
and $\chi_{ns}$ can be chosen as 
\begin{eqnarray}
   && \hspace{-1em}
   \chi_{n,1}(p_z,y) = {a_{\rm m}^{-1/2}\over\sqrt{2E(E+1)}}\left(
   \begin{array}{c}
      (E+1)\,{\cal H}_{n-1}(y/a_{\rm m}) \\
      0 \\
      P_{\eta n}(E)\,{\cal H}_{n-1}(y/a_{\rm m}) \\
      -\sqrt{2bn}\,{\cal H}_n(y/a_{\rm m})
   \end{array}
   \right),
\nonumber \\
   && \hspace{-1em}
   \chi_{n,-1}(p_z,y) \! =\! {a_{\rm m}^{-1/2}\over\sqrt{2E(E+1)}}
   \! \left(
   \begin{array}{c}
      0 \\
     (E+1)\,{\cal H}_n(y/a_{\rm m}) \\
     -\sqrt{2bn}\,{\cal H}_{n-1}(y/a_{\rm m}) \\
     -P_{\eta n}(E)\,{\cal H}_n(y/a_{\rm m})
   \end{array}
   \right)\! .
\label{3.2}
\end{eqnarray}
Here $b=\hbar\omega_B/(m_{\rm e}c^2)$,
$E = \epsilon/(m_{\rm e}c^2)$ and
$P_{\eta n}(E) = \eta\, (E^2-1-2bn)^{1/2}$
are, respectively, the magnetic field, the energy
and the longitudinal momentum in the relativistic units,
$\eta={\rm sign}\,p_z$,
\begin{equation}
   \,{\cal H}_n(\xi)= {\exp(-\xi^2/2)\over \pi^{1/4}(2^n n!)^{1/2}}
   H_n(\xi),
\label{3.3}
\end{equation}
and $H_n(\xi)$ is a Hermite polynomial. 

In the non-relativistic limit, the two lower components of
the bispinors (\ref{3.2}) are negligible
and this basis corresponds to fixed
spin projections ($s\hbar/2)$ on the $z$-axis. However one can
use another basis: 
\begin{equation}
   \chi'_{ns} = \chi_{ns}\cos\phi - s\, \chi_{n,-s}\sin\phi\,.
\label{3.4}
\end{equation}
It is sufficient to assume that
$0\leq\phi\leq\pi/2$; $\phi$ may depend on $n$ and
should vanish for $n=0$. In particular, the choice
\begin{equation}
    \phi = \phi_n \equiv
    \arcsin\left({1\over 2}\left[1-(E^2-1)^{-1/2}P_{\eta n}(E) \right]
    \right)^{1/2}
\label{3.5}
\end{equation}
yields the basis of states with fixed helicity
used by Hernquist (1984) and Schaaf (1988).

Using the distribution function formalism,
one generally obtains different conductivities
with different basic functions.
Yakovlev (1984) employed both $\phi=0$ and $\phi=\phi_n$ 
and interpolated between corresponding results assuming
that the former choice is appropriate in the non-relativistic limit
($E\sim 1$) while the latter one is adequate in the
ultrarelativistic case ($E\gg 1$).

The accuracy of the distribution function calculations 
can be verified using the density matrix, 
\begin{equation}
   \rho_{\kappa'\kappa} =
   \langle\psi_{\kappa'}|\hat{\rho}|\psi_\kappa\rangle,
\label{3.6}
\end{equation}
where $\hat{\rho}$ is the statistical operator.

Let us consider elastic collisions
and assume that the density matrix is diagonal in energy:
$\epsilon'=\epsilon$. Furthermore, we assume the diagonality
in momenta. We shall show that the latter property is not violated by
collisions. The diagonalities in $\epsilon$ and $p_z$ lead to 
the diagonality in $n$. The diagonality of the density matrix
in $p_x$, $p_z$ allows us to treat
the dependence of $\rho_{\kappa'\kappa}$ on $x$, $z$ parametrically,
without using the Wigner transformation.
Thus, we can write
\begin{equation}
   \rho_{\kappa'\kappa} = \delta_{n'n}\,
   \delta_{p'_x p_x}\,\delta_{p'_z p_z}\,
   \rho_{ns's}(x,z,p_x,p_z), 
\label{3.7}
\end{equation}
where
$   \delta_{p'_\alpha p_\alpha} 
\equiv (2\pi\hbar/ L_\alpha)\,\delta(p'_\alpha-p_\alpha).$ 
Analysing the longitudinal transport properties,
we can assume that $\hat{\rho}$ is independent of $x$.
Let us introduce the spin-polarization
density matrix,
\begin{equation}
   \rho_{\eta n s' s}(z,\epsilon) =
   \int \rho_{n s's}(z,p_x,p_z)\,{L_x\over 2\pi\hbar}\,{\rm d}p_x.
\label{3.8}
\end{equation}
Then we obtain the kinetic equation
with the classical left-hand side
\begin{equation}
   \left( {\partial\over\partial t}+v_z{\partial\over\partial z}-
   e{\cal E}{\partial\over\partial p_z}\right)
   \rho_{\eta n s_1 s_2}(z,\epsilon) =
   \left[{{\rm d}\rho_{\eta n s_1 s_2}\over {\rm d}t}\right]_{\rm c},
\label{3.9}
\end{equation}
where ${\cal E}$ is a longitudinal electric field and $v_z$ 
is the velocity. The density matrix depends 
on $p_z$ through $\epsilon$ and $\eta$. The right-hand side
of Eq.\,(\ref{3.9}) is the collision integral to be
determined from microscopic considerations. Owing to the linearity
of the equations which govern the electron wave-function
evolution during a scattering event,
the collision integral should be linear in $\rho_{\eta n s_1 s_2}$. 
Thus it can be presented as 
\begin{equation}
   \left[{{\rm d}\rho_{\eta n s_1 s_2}\over {\rm d}t}\right]_{\rm c} 
   =
   \sum_{\eta' n' s'_1 s'_2} 
   D_{\eta' n' s'_1 s'_2; \eta n s_1 s_2}\,
   \rho_{\eta'n's'_1 s'_2}.
\label{3.10}
\end{equation}
The equilibrium density matrix is 
$\rho^{(0)}_{s_1 s_2}(\epsilon) = 
f_0(\epsilon)\,\delta_{s_1 s_2}$, and the collision
integral vanishes at the equilibrium.

Deviations from the equilibrium in the linear regime
can be treated as small perturbations. In the zero-order
approximation, the density matrix is equal to $\rho^{(0)}$; it
depends on $z$ parametrically through $\mu$ and $T$.
In the first-order approximation, it is customary to write 
\begin{eqnarray}
   \rho_{\eta n s_1 s_2}(z,\epsilon) & = &
   \rho^{(0)}_{s_1 s_2}(\epsilon) +
   \eta l\,
   {\partial f_0(\epsilon)\over \partial\epsilon}\,
   \left[e{\cal E}+{\partial\mu\over\partial z}+
   \right.
\nonumber \\
   && \left.
   {\epsilon - \mu \over T}\,{\partial T\over\partial z}
   \right]\,\varphi_{\eta n s_1 s_2}(\epsilon),
\label{3.12}
\end{eqnarray}
where $l$ is the scale length defined by Eq.\,(\ref{2.8}).
Keeping the zero-order terms on the
left-hand side of Eq.\,(\ref{3.9}) we obtain the set of
algebraic equations for the non-equilibrium corrections $\varphi$:
\begin{equation}
   -{l\over|v_z|}\!\sum_{\eta'n's'_1 s'_2}\!\! \eta'
   D_{\eta'n's'_1 s'_2;\eta n s_1 s_2}\,
   \varphi_{\eta' n' s'_1 s'_2} 
   = \eta\, \delta_{s_1 s_2}. 
\label{3.13}
\end{equation}
A solution of these equations allows us to calculate
the kinetic coefficients $\sigma$, $\beta$, and $\lambda$
in the transport relations (\ref{3.14}) and so determine 
the longitudinal electric and thermal conductivities
and thermopower.

Using Eq.\,(\ref{3.12}) we can write $\sigma$, $\beta$ and $\lambda$
in the form of Eq.\,(\ref{3.15}),
with the effective relaxation time defined as
\begin{eqnarray}
   && \hspace{-1em}
   \tau(\epsilon) =
   {\epsilon l m_{\rm e}
   \omega_B\over 2(\pi\hbar c)^2n_{\rm e}}\,\Phi(\epsilon),
\label{4.1}
\\
   && \hspace{-1em}
   \Phi(\epsilon) = {1\over 2}\sum_{\eta=\pm1}
   \sum_{n=0}^{n_\epsilon} \sum_{s=\pm1}
   \varphi_{\eta nss}(\epsilon).
\label{3.16}
\end{eqnarray}
Here $n_\epsilon= {\rm Int}(\nu)$ specifies the highest Landau level
populated by electrons with energy $\epsilon$, and 
\begin{equation}
   \nu = {\epsilon-m_{\rm e}c^2 \over \hbar\omega_B}\,
   {\epsilon+m_{\rm e}c^2 \over 2m_{\rm e}c^2} 
   = (E^2 - 1)/(2b).
\label{3.17}
\end{equation}
Equations (\ref{3.15}) and (\ref{4.1}) reproduce 
Eq.\,(24) of Yakovlev (1984). However now
Eq.\,(\ref{3.16}) contains diagonal elements of the density matrix 
(to be determined from Eq.\,(\ref{3.13}))
instead of the distribution function in the traditional approach.
The factors $D$ in Eq.\,(\ref{3.13}) are obtained in
Sect. \ref{sect3.2}. They are found to be
much more complicated than analogous factors
in the traditional equations.

\subsection{Collision integral}                          
\label{sect3.2}
First let us derive the collision integral for the
full density matrix $\rho_{\kappa'\kappa}$.
After that the collision integral (\ref{3.10}) is obtained
by summing over those quantum numbers in which $\rho_{\kappa'\kappa}$
is diagonal.

In the ``quasiclassical approach'' for the density 
matrix (Sobelman et al. 1981), interaction of electrons 
with perturbers (ions or phonons) is described by a scattering
potential $V_0(t)$. 
A wave function of an interacting electron evolves as
$\psi(t)=\hat{S}(t,t_0)\psi(t_0)$, where $\hat{S}$ is
the scattering operator. Substituting this into Eq.\,(\ref{3.6}) 
and averaging over collision parameters we obtain
\begin{equation}
   \rho_{\kappa_1\kappa_2}(t) =
   \sum_{\kappa'_1\kappa'_2} \langle
   S^\ast_{\kappa'_1\kappa_1}(t)
   S_{\kappa'_2\kappa_2}(t) \rangle_{\rm pc} 
   \rho_{\kappa'_1\kappa'_2}(-\infty), 
\label{3.18}
\end{equation}
where 
$    
S_{\kappa'\kappa}\equiv 
\langle\kappa'|\hat{S}(t,-\infty)|\kappa\rangle 
$ 
is the scattering matrix, and 
the brackets $\langle\ldots\rangle_{\rm pc}$ 
denote space averaging over perturber
centers ${\bf r}_0$ (i.e., space integration with the weight 
$n_{\rm i}$), 
while the brackets without subscript denote the quantum-mechanical 
averaging. 
The sum over $\kappa'$ includes that over
$p'_x$ and $p'_z$, which should be performed according
to the correspondence rule
\begin{equation}
   \sum_{p_{x,z}} \longleftrightarrow
   \int {L_{x,z}\over 2\pi\hbar}\,{\rm d}p_{x,z}.
\label{3.19}
\end{equation}
Following the ``adiabatic switch on''
method (Landau \& Lifshitz 1976), we put 
\begin{equation}
   V_0(t) = V({\bf r} - {\bf r}_0)\,{\rm e}^{\varepsilon t}, 
\label{3.20}
\end{equation}
where $\varepsilon \rightarrow 0$ is the adiabatic parameter, 
and $V({\bf r})$ is the actual scattering potential. Now we obtain 
\begin{equation}
   \left[{{\rm d}\rho_{\kappa_1\kappa_2}\over {\rm d}t}\right]_{\rm c} =
   \sum_{\kappa'_1\kappa'_2} D_{\kappa'_1\kappa'_2 \kappa_1\kappa_2}\,
   \rho_{\kappa'_1\kappa'_2},
\label{3.21}
\end{equation}
where
\begin{equation}
   D_{\kappa'_1\kappa'_2 \kappa_1\kappa_2} =
   \lim_{\varepsilon\to 0}\left<\left.
   {{\rm d}\over {\rm d}t}
   \left[S^\ast_{\kappa'_1\kappa_1}(t)
   S_{\kappa'_2\kappa_2}(t) \right]
   \right|_{t=0}\right>_{\rm pc} . 
\label{3.22}
\end{equation}
The scattering matrix elements $S_{\kappa'\kappa}$
can be calculated in the second-order perturbation
theory (e.g., Landau \& Lifshitz 1976).
Let us substitute them
into Eqs.\,(\ref{3.22}) and (\ref{3.21}).
Then the terms linear in $V$ are canceled
after averaging over ${\bf r}_0$, while the quadratic terms give 
\begin{eqnarray}
   && \hspace{-1em}
   D_{\kappa'_1\kappa'_2, \kappa_1\kappa_2}  =
   -{\pi\over\hbar^2}\sum_\kappa
   \left[\delta_{\kappa'_2\kappa_2} \delta(\omega_{\kappa\kappa_1})
   \langle V^\ast_{\kappa\kappa_1}V_{\kappa\kappa'_1}\rangle_{\rm pc} +
   \right.
\nonumber \\ &&
~~~~~~
   \left.
   \delta_{\kappa'_1\kappa_1}\delta(\omega_{\kappa\kappa_2})
   \langle V^\ast_{\kappa\kappa'_2}V_{\kappa\kappa_2}\rangle_{\rm pc}
   \right] -
\nonumber \\ &&
   ~~~~~~
   {{\rm i}\over\hbar^2}\sum_\kappa \left[
   \delta_{\kappa'_2\kappa_2}
   {\cal P}{1\over\omega_{\kappa\kappa_1}}
   \langle V^\ast_{\kappa\kappa_1}V_{\kappa\kappa'_1}\rangle_{\rm pc} -
   \right.
\nonumber \\ &&
   ~~~~~~
   \left.
   \delta_{\kappa'_1\kappa_1}
   {\cal P}{1\over\omega_{\kappa\kappa_2}}
   \langle V^\ast_{\kappa\kappa'_2}V_{\kappa\kappa_2}\rangle_{\rm pc}
   \right] +
\nonumber \\ &&
   ~~~~~~
   {\pi\over\hbar^2}\langle
   V^\ast_{\kappa'_1\kappa_1}V_{\kappa'_2\kappa_2}\rangle_{\rm pc}
   \left(\delta(\omega_{\kappa'_1\kappa_1})+
   \delta(\omega_{\kappa'_2\kappa_2})\right) +
\nonumber \\ &&
   ~~~~~~
   {{\rm i}\over\hbar^2}\langle
   V^\ast_{\kappa'_1\kappa_1}V_{\kappa'_2\kappa_2}\rangle_{\rm pc}
   \left({\cal P}{1\over\omega_{\kappa'_1\kappa_1}} -
   {\cal P}{1\over\omega_{\kappa'_2\kappa_2}}
   \right),
\label{3.24}
\end{eqnarray}
where $\omega_{\kappa'\kappa}\equiv (\epsilon'-\epsilon)/\hbar$, 
and $V_{\kappa'\kappa}$ is the matrix element of the potential $V$.
In deriving Eq.\,(\ref{3.24}) we have taken 
into account the well known relationship 
\begin{equation}
   \lim_{\varepsilon\to 0}{1\over\varepsilon\mp{\rm i}\omega} =
   \pi\delta(\omega)\pm {\rm i}{\cal P}{1\over\omega},
\label{3.23}
\end{equation}
where ${\cal P}$ denotes the Cauchy main part. 
An averaged binary product in Eq.\,(\ref{3.24}) is equal to
\begin{eqnarray}
   & & \hspace{-1em}
   \langle
   V^\ast_{\kappa'_1\kappa_1}V_{\kappa'_2\kappa_2}\rangle_{\rm pc} =
   {2\pi n_{\rm i}\over (L_x L_z)^2}\,\delta(q_{1x}-q_{2x})
   \delta(q_{1z}-q_{2z}) \!\!\int \!\!{\rm d}q_y\!\!
   \times
\nonumber \\ &&\hspace{-1em}
   |U({\bf q})|^2
   M^\ast_{s'_1 s_1}(n',p'_z; n_1, p_{1z}; u)
   M_{s'_2 s_2}(n',p'_z; n_2, p_{2z}; u),
\label{3.25}
\end{eqnarray}
where ${\bf q}\equiv ({\bf p'}-{\bf p})/\hbar$, 
$U({\bf q})$ is the Fourier image of the potential,
and
\begin{eqnarray}
   & & \hspace{-1em}
   M_{s's}(n',p'_z;n,p_z;u) =
\nonumber \\ &&\hspace{-1em}
   \int_{-\infty}^{\infty}\!\!\!
   \chi^+_{n's'}(p'_z, y - q_x a_{\rm m}^2/2) \chi_{ns}(p_z, y+q_x 
   a_{\rm m}^2/2)
   {\rm e}^{{\rm i}q_yy}{\rm d}y
\label{3.26}
\end{eqnarray}
is the matrix element which actually depends on $q_x$ and $q_y$ only
through the variable $u\equiv (q^2_x+q^2_y)a^2_{\rm m}/2$
(e.g., Kaminker \& Yakovlev 1981).
In Eq.\,(\ref{3.25}) we have set
$p'_{1x}=p'_{2x}$, $p'_{1z}=p'_{2z}$, and 
$n'_1=n'_2$, since the density matrix is diagonal in these
arguments. Due to the same diagonality, 
$\omega_{\kappa'_1\kappa_1} = \omega_{\kappa'_2\kappa_2}$, 
and the last term in Eq.\,(\ref{3.24}) vanishes.

Finally, let us sum $D_{\kappa'_1\kappa'_2 \kappa_1\kappa_2}$
over $n_2$, $p_{1x}$, $p_{2x}$, and $p_{2z}$ using
Eq.\,(\ref{3.19}). Thus we arrive at the collision integral
(\ref{3.10}), where
\begin{eqnarray}
   & & \hspace{-1em}
   D_{\eta'n's'_1 s'_2;\eta n s_1 s_2} =
   {|v_z| \over l}\, [a_{\eta n s_1 s_2, \eta'n's'_1 s'_2} -
\nonumber \\
   && {1\over 2}\,\delta_{\eta'\eta}\delta_{n'n}\,
   (A_{\eta n s_1 s'}\delta_{s'_1 s'}\delta_{s'_2 s_2} +
   A^\ast_{\eta n s's_2}\delta_{s'_1 s_1}\delta_{s's_2})].
\label{3.27}
\end{eqnarray}
Explicit expressions for the real coefficients $a$ and complex 
coefficients $A$ depend on the basis. In the basis (\ref{3.2}), 
they are given in Appendix~A. 

\subsection{Algebraic equations for the density matrix}     
\label{sect3.3}
Transformation properties of the coefficients $a$ and $A$
with respect to the reflection $p_z\to -p_z$ or $p'_z\to -p'_z$ 
allow us to split the system (\ref{3.13}) into two
independent subsystems for $\eta=1$ and $\eta=-1$.
Using the basis (\ref{3.2}), we have 
\begin{equation}
   \varphi_{\eta nss} = \varphi_{-\eta,nss}
   ~~ {\rm and} ~~
   \varphi_{\eta ns,-s} = -\varphi_{-\eta,ns,-s}.
\label{3.28}
\end{equation}
Thus it is sufficient to solve one of the subsystems: 
\begin{eqnarray}
   {1\over 2}\sum_{s'}\left(A_{ns_1s'}\varphi_{ns's_2} +
   A^\ast_{ns's_2}\varphi_{ns_1s'} \right)
   & - &
\nonumber \\[-1ex]
   \sum_{\eta'n's's''} \gamma a_{ns_1s_2,\eta'n's's''}\,
   \varphi_{n's's''} & = & \delta_{s_1s_2} ,
\label{3.29}
\end{eqnarray}
where 
$\gamma=\eta'$ for $s' = s''$, and $\gamma=1$ otherwise.
In Eq.\,(\ref{3.29}) we have set
$\eta=1$ in $a$, $A$, and $\varphi$.

Retaining diagonal elements of $\varphi$ and
setting $s_1=s_2$ in
Eq.\,(\ref{3.29}) we recover the algebraic equations
derived by Yakovlev (1984) in the
distribution function formalism. However thus reduced system is not
covariant with respect to the basis transformations (\ref{3.4})
(note that the coefficients $a$ and $A$ have been obtained from the
matrix elements (\ref{3.26}), and they undergo
the transformations together with
the density matrix elements).
The lack of covariance leads to the 
dependence of the electron transport coefficients on 
the basis. For instance, Yakovlev (1984) obtained the
difference up to 20\% using different basis sets
for some particular electron gas parameters.
Our density matrix formalism makes the complete
system (\ref{3.13}) covariant with
respect to the basis transformations.
The trace of the density matrix is invariant, 
therefore changing the basis does no more 
affect $\Phi(\epsilon)$
(Eq.\,(\ref{3.16})) and the kinetic coefficients.

\subsection{Applicability range}                          
\label{sect3.4}
Let us discuss briefly the validity of the assumptions
which led us to the algebraic system (\ref{3.29}),
for the neutron star crust conditions.
First, we have assumed diagonality of the density matrix
in $p_x$ and $p_z$. This property is
practically exact. Indeed, the delta-functions
$\delta(q_{1x}-q_{2x})\,\delta(q_{1z}-q_{2z})$ 
in Eq.\,(\ref{3.25}) enter the
right-hand side of Eq.\,(\ref{3.21}) through Eq.\,(\ref{3.24}).
Since $\rho_{\kappa'_1\kappa'_2}$
is diagonal in $p'_x$ and $p'_z$, the
right-hand side of Eq.\,(\ref{3.21}) 
virtually contains 
$\delta(p_{1x}-p_{2x})\,\delta(p_{1z}-p_{2z})$.
Thus the diagonality of the
density matrix in $p_x$ and $p_z$ is not affected by collisions.
Note that the delta-functions in
Eq.\,(\ref{3.25}) appeared due to the infinitely large
volume assumed in averaging over ${\bf r}_0$.
If the volume were finite, a non-diagonality in the momenta occurred
in the band $\Delta p\sim\hbar/L$, in agreement with the uncertainty
principle. An actual value of $L$ is restricted by the 
condition of spatial uniformity of the considered bulk of matter. 
   The broadening $\Delta p$ is negligible compared with 
   the typical momentum difference between adjacent Landau orbitals, 
   provided that $L\gg \hbar/(m_{\rm e}c\sqrt{b})$, which condition 
   is always satisfied in the magnetized neutron star crust. 

Another assumption is concerned with the diagonality 
in the Landau numbers.
The last term in Eq.\,(\ref{3.24}) shows that this property
is not conserved in collisions. This effect can be estimated 
with the aid of the uncertainty principle.
The non-diagonality should occur within
the collisional band $\Delta\epsilon\sim\hbar/\tau$, 
where $\tau$ is the effective relaxation time. The density matrix 
is diagonal in $n$ if $\Delta\epsilon < \hbar\omega_B^\ast$. The electron
cyclotron energy $\hbar\omega_B$ ranges from $10^2$ to $10^4$~Ry
(where Ry =13.6~eV is the Rydberg energy),
for magnetic fields from $10^{11}$ to $10^{13}$~G.
For estimating $\Delta\epsilon$, we 
can use the non-magnetic relaxation times
(\ref{4.10b}), (\ref{4.10c}) (cf. Yakovlev 1984). 
If $Z\sim 26$, $T \sim 10^7$~K, and
the electrons are mildly relativistic,
we obtain $\hbar/\tau \sim 
\Delta\epsilon\sim 30~{\rm Ry}~ < \hbar\omega_B$. 
This estimate relates to the collisional broadening. 
Other broadening mechanisms, not treated here, are 
due to non-elasticity of scattering and deviations from 
the Born approximation. 
Yakovlev (1984) argued that the two latter types 
of broadening seem to be unimportant under the considered 
physical conditions. 

While considering the spin number $s=\pm1$, one may notice that the 
spin-polarization density matrix $\rho_{\eta n s' s}$ 
is reduced compared to the general case of $4\times4$ matrix 
(Berestetskij et al. 1982). It is because we restrict the 
basis to the electron bispinors $\chi_{ns}$, thus 
neglecting an admixture of positron states. 
Of course, this restriction is well 
justified at the non-relativistic temperatures of interest. 

Finally, the assumed degeneracy in the spin number 
is not exact, owing to the quantum-electrodynamical corrections. 
The corrections split each Landau level into two 
sublevels (e.g., Landau \& Lifshitz 1982). 
According to the Schwinger formula, the splitting energy is about 
$(\alpha/2\pi)\hbar\omega_B\sim (0.1 - 10)~{\rm Ry}~ < \Delta\epsilon$,
for $B=10^{11} - 10^{13}$~G. 
Therefore the collisional width of the sublevels exceeds 
their separation, and they can be considered as degenerate. 

Thus Eq.\,(\ref{3.29}) is valid
for the typical conditions in neutron star crusts.

\subsection{Numerical examples and discussion}              
\label{sect3.5}
We have computed the first-order correction 
$\varphi_{ns_1s_2}(\epsilon)$ to the density matrix
and its trace $\Phi(\epsilon)$ for a wide range of magnetic fields
from $10^{11}$ to $10^{14}$~G.
The coefficients in Eq.\,(\ref{3.29})
have been calculated using the formulae from
Appendices A and B.
The system (\ref{3.29}) has been solved using the
$LU$ factorization code by Fletcher (1988). We have calculated
the density matrix and also the electron distribution function
in two representations, with $\phi=0$ and $\phi=\phi_n$ 
(Eqs.\,(\ref{3.2}), (\ref{3.4}), and (\ref{3.5})).
The coefficients of the algebraic systems 
for the distribution functions have been taken from Yakovlev (1984),
and the integrals $Q_i$
which enter these coefficients have been calculated 
using the formulae of Appendix~B.

At relatively low field strengths, the
density matrix formalism gives
practically the same results as
the distribution function approach with
the basis (\ref{3.2}) ($\phi=0$),
in agreement with anticipation of Yakovlev (1984).
The inaccuracy of the latter approach is about 0.1\% for
$B\la 10^{12}$~G.
The distribution function results based on the alternative
``fixed helicity'' representation
($\phi=\phi_n$), employed by Hernquist (1984) and Schaaf (1988),
deviate from the exact results by several per cent.
When $n_\epsilon=1$, 
the deviations reach 3\% for the phonon scattering
and 8\% for the Coulomb scattering. 
The deviations decrease at higher $n_\epsilon$.

The discrepancies between different representations are larger
for stronger fields.
Figure~\ref{fig1} shows the results for the iron crust ($Z=26$) 
with magnetic fields 
$B=10^{13}$ and $10^{14}$~G. The upper panels display
the function $\Phi(\epsilon)$ for the electron energies
sufficient to occupy up to ten Landau levels.
One can see strong quantum oscillations of $\Phi(\epsilon)$. 
Sharp dips at integer values of $\nu$ are caused by 
singularities of the electron density of states due to the 
magnetic quantization, when electrons start to populate 
new Landau levels. 

The lower panels show relative errors of the 
distribution function approximation 
for two basis choices discussed above. 
The discrepancies reach up to 20\% at $B=10^{14}$~G for 
the Coulomb scattering, in agreement with results of Yakovlev (1984). 
However, the energy dependence of the relative errors 
is non-monotonous, and a simple interpolation between 
the approximate results proposed by Yakovlev (1984) 
is not very accurate. 
The basis (\ref{3.2}) provides higher accuracy in most cases, but 
the alternative basis (\ref{3.4}), (\ref{3.5}) 
becomes more appropriate at relativistic 
energies for phonon scattering.

Generally, the discrepancies between different representations
decrease with increasing energy. We have checked numerically,
that relative errors do not exceed 1\% for $\nu>50$ even
at $B=10^{14}$~G. Calculations for such high $\nu$
are facilitated by semiclassical formulae for the integrals
$Q_i$ derived in Appendix~B.

Our calculations show that one should use the density matrix
formalism for energies $1<\nu \la 5$ and relativistic
magnetic fields $b \ga 1$, if one 
needs to keep an error within a few percent.
Otherwise the distribution function approach 
with the basis (\ref{3.2}) 
provides the desired accuracy. 

For a better comparison with the results of Yakovlev (1984) and
Hernquist (1984), the examples in
Fig.\,\ref{fig1} have been calculated with
the scattering potentials used by these
authors. In particular,
the screening length in the liquid regime
has been determined by Eqs.\,(\ref{2.2c}) and (\ref{2.2d}). 
In the solid regime, we have adopted
the conditions when the Debye -- Waller factor can be
neglected. Under these assumptions, $\Phi(\epsilon)$
does not depend on temperature. The effect of finite
temperatures on $\Phi(\epsilon)$ is generally small in the liquid 
regime. For the solid regime, the effect is discussed in the
next section.

\section{High-temperature conductivity enhancement 
in the solid phase}                                          
\label{sect4}
As seen from Eqs.\,(\ref{2.3}) -- (\ref{2.4a}), 
the Debye -- Waller reduction of electron-phonon scattering 
becomes important at sufficiently high 
temperatures. Its influence on the electron transport in dense 
non-magnetized stellar matter has been 
considered by Itoh et al. (1984a, 1993). 
The effect is most pronounced near the melting point.
In this case it can be easily estimated. Let us put
$p=p_0$, where $p_0=\hbar(3\pi^2 n_{\rm e})^{1/3}$ is the field-free
Fermi momentum. Taking into account Eq.\,(\ref{2.4}), we can
write $w$ in Eq.\,(\ref{4.10a}) as
$w=(4/3)u_{-2}(9\pi Z/4)^{2/3}/\Gamma$, which gives 
$w=560/\Gamma$ for $u_{-2}=13$, $Z=26$. Near the melting point
$\Gamma\sim 180$, and $w\sim 3$. Then the
reduction factors are $R_1\sim 0.3$ and $R_2\sim 0.2$.
According to Eq.\,(\ref{4.10c}), this leads to the largest
increase of the relaxation time by a factor of 3.3 for
non-relativistic electrons and 2.5 for ultrarelativistic 
electrons, in qualitative agreement with numerical 
results of Itoh et al. (1984a, 1993). 

However the Debye -- Waller factor has been neglected so far
while studying the electron transport in quantizing magnetic
fields (Yakovlev 1984, Hernquist 1984, Schaaf 1988, Van Riper 
1988). In this case, contrary to the non-magnetic one, 
the longitudinal momentum $p_z$ is quantized into the discrete values 
$mc\eta P_{\eta n}$ (see Sect.\,\ref{sect3.1}), 
and only scattering events 
with changing $p_z$ do contribute to the transport processes. 
Therefore there exists a lowest collisional momentum transfer 
$\Delta p_{\rm min}(\epsilon)$. 
According to Eqs.\,(\ref{A11}) and (\ref{A12}) 
of Appendix~A,
the scattering rate acquires an additional exponential
reduction argument
$\zeta\xi_{\rm min} = (r_T\Delta p_{\rm min}/\hbar)^2/3$. 
The smallest momentum difference reaches 
$\Delta p_{\rm min} \sim 2\sqrt{2m_{\rm e}\hbar\omega_B}$ 
just below the first Landau threshold, and 
$\Delta p_{\rm min} \sim m_{\rm e}\hbar\omega_B/p_0$ 
just before each next threshold. 
Therefore the exponent argument is
$\zeta\xi_{\rm min} \sim w$ for $n_\epsilon=0$, and 
$\zeta\xi_{\rm min} \sim w(m_{\rm e}\hbar\omega_B/p_0^2)^2$ 
for high $n_\epsilon$. In the latter case, the 
additional exponent argument becomes small and unimportant. 
However, it is significant at low $n_\epsilon$.

Figure~\ref{fig2} shows the Debye -- Waller
enhancement of the electron relaxation
time in quantizing magnetic
fields. For each $T$, $B$, and electron energy $\epsilon$
we have put $\mu=\epsilon$ and 
calculated the electron number density from
Eq.\,({\ref{2.0}) to determine the Debye -- Waller exponent
argument. The fitting formulae presented in Appendix~C 
facilitate this calculation. 
Then the coefficients of the algebraic system (\ref{3.29}) 
were computed, using the formulae of Appendices A, B, 
and the system was solved to obtain the function 
$\Phi(\epsilon)$,  which is proportional to 
the effective relaxation time $\tau(\epsilon)$, Eq.\,(\ref{3.16}).
Its ratio to the same function
calculated without the Debye -- Waller factor is plotted
against the Landau variable $\nu$ in Fig.~\ref{fig2}.  
This ratio reveals quantum oscillations in phase with 
oscillations of the function $\Phi(\epsilon)$ itself, thus 
increasing the amplitude of the latter ones. For comparison, 
the non-magnetic enhancement factor $(R_1-(R_2/2)(v/c)^2)^{-1}$ 
(cf.  Eq.\,(\ref{4.10c})) is shown by dashes. As seen from 
Fig.~\ref{fig2}, the quantizing field makes the Debye -- Waller
enhancement much stronger. When the electrons populate 
one or two Landau levels, the relaxation 
time is enhanced by a factor of up to 30 
(compared with the factor of about 3 in the non-magnetic case).
\section{Summary}                                             
\label{sect5}
We have presented the theory of transport properties of
degenerate electrons along quantizing magnetic fields
in neutron star crusts. Our results are advanced, 
compared to the previously known ones, in three 
respects. 

First, a kinetic equation for the spin polarization density matrix 
of electrons is derived. The solution of this 
equation provides a justification of 
the standard approach based on the kinetic equation 
for the electron distribution function. The present results are 
compared with two versions of the standard approach used 
previously by different authors. 

For non-relativistic magnetic
fields, $B\la 10^{13}$~G, our results confirm the arguments
of Yakovlev (1984) that the standard approach
which employs basic functions with fixed spin
$z$-projection is the most appropriate in the non-relativistic
limit. The fixed-helicity basic functions used by other authors
lead to small inaccuracies which however seem to be
insignificant in astrophysical implications.

For stronger fields, $B\ga 10^{13}$~G, the inaccuracies
of the traditional approach increase up to 20\% when density
is rather low and
the electrons occupy low-lying Landau levels. 
The density-matrix results allow us 
to choose the most appropriate version of the standard approach.
If density is higher and the electrons populate many Landau levels,
the difference between various approaches becomes negligible.

Secondly, we have taken into account
the cumulative effect of the Debye -- Waller factor with the magnetic 
quantization. In the non-magnetic case, this
factor can increase the thermal and electric conductivities by
a factor of 3 just below the melting temperature. We
show that the magnetic quantization can enhance the
effect by an order of magnitude.

Thirdly, we have derived semiclassical expressions
for some intermediate integrals
which enter the system of equations either for the
density matrix or for a distribution function. These expressions
provide fast and accurate calculation of the relaxation time for
large number of occupied Landau levels. 

In this paper we have calculated the
kernel function $\Phi$ which
should undergo further thermal averaging, Eq.\,(\ref{3.15}),
to determine the longitudinal electric and thermal conductivities
and thermopower. We shall consider
this averaging and astrophysical
implications of the developed theory in the subsequent paper.

\acknowledgements 
I am greatly indebted to D.G.\,Yakovlev for introducing me 
in this problem and numerous helpful discussions. Also 
I am pleased to acknowledge useful discussions with 
C.J.\,Pethick, D.\,Page, and G.G.\,Pavlov, 
and valuable comments by K.\,Van Riper. 
I am grateful to all Nordita staff 
and especially to C.J.\,Pethick for the hospitality 
during my stay at Nordita. This work was 
supported in part 
by the Nordic Council of Ministers' Scholarship Programme 
for the Baltic Region and North-West Russia, 
by the International Science Foundation 
(grant R6A000), 
and by the ESO E\&EC Programme (grant A-01-068).

\appendix
\addtocounter{section}{1}
\section*{Appendix A: coefficients in equations for the
density matrix}                                          
While deriving Eq.\,(\ref{3.27}), let us first reduce Eq.\,(\ref{3.21})
to Eq.\,(\ref{3.10}). For this purpose,
let us sum Eq.\,(\ref{3.21}) over
$p_{2x}$, $p_{2z}$, $n_2$ and $p_{1z}$ taking into account
Eqs.\,(\ref{3.7}), (\ref{3.8}), (\ref{3.24}) -- (\ref{3.26}).
This yields
\begin{eqnarray}
   && \hspace{-1em}
   \left[{{\rm d}\rho_{\eta n s_1 s_2}\over {\rm d}t}\right]_{\rm c} =
   -\,{\pi\over\hbar^2}\sum_{\eta'n'p'_zs's''}
   \left(K_{s's_1,s's''}\rho_{\eta n s''s_2}(p_z) +
   \right.
\nonumber \\ &&
   \left.
   K_{s's'',s's_2}\rho_{\eta n s_1 s''}(p_z) \right)\,\delta(\omega) -
\nonumber \\ &&
   {{\rm i}\over\hbar^2}\sum_{\eta'n'p'_zs's''}
   \left(K_{s's_1,s's''}\rho_{\eta n s''s_2}(p_z) -
\right.
\nonumber \\ &&
\left. 
   K_{s's'',s's_2}\rho_{\eta n s_1 s''}(p_z) \right)\,
   {\cal P}{1\over\omega} +
\nonumber \\ &&
   {2\pi\over\hbar^2} \sum_{\eta'n'p'_zs's''}
   K_{s's_1, s''s_2}\rho_{\eta'n's's''}(p'_z)\,\delta(\omega),
\label{A1}
\end{eqnarray}
where $\omega = (\epsilon'-\epsilon)/\hbar$ and 
\begin{eqnarray}
 && \hspace{-1em}
   K_{s's_1,s''s_2}  \equiv 
   K_{s's_1,s''s_2}(n', p'_z; n, p_z) =
   {n_{\rm i}\over (2\pi)^2L_z}\int {\rm d}q_x{\rm d}q_y\,\times 
\nonumber \\ && 
   |U({\bf q})|^2 
   M^\ast_{s' s_1}(n',p'_z; n, p_z; u)
   M_{s'' s_2}(n',p'_z; n, p_z; u) .
\label{A2}
\end{eqnarray}
It is convenient to introduce new integration variable
$u = (q_x^2+q_y^2)\,a_{\rm m}^2/2$
and the dimensionless function
\begin{equation}
   v(u,\xi) = \left({n_{\rm i} l\over 2\pi}\right)^{1/2}(\hbar c 
   a_{\rm m})^{-1}
   |U({\bf q})|.
\label{A3}
\end{equation}
For the Coulomb scattering, 
\begin{equation}
   v(u,\xi) = (u+\xi)^{-1}, ~~~ \xi = (q_z^2 + r_{\rm s}^{-2})\,
   a_{\rm m}^2/2,
\label{A4}
\end{equation}
where $r_{\rm s}$ is the screening length, and $\hbar q_z = p'_z-p_z$. 
For the scattering on phonons, 
\begin{equation}
   v(u,\xi) = (u+\xi)^{-1/2}{\rm e}^{-W}, 
   ~~~ \xi = q_z^2 a_{\rm m}^2/2.
\label{A5}
\end{equation}
The sum over $p'_z$ in Eq.\,(\ref{A1}) can be converted into 
the sum over $\eta'$ and the integral over $\epsilon'$. 
The delta-function $\delta(\omega)$ eliminates the integration
in the first and the last terms of Eq.\,(\ref{A1}).
Finally we arrive at Eq.\,(\ref{3.27}) with
\begin{eqnarray}
   && \hspace{-1em}
   a_{\eta n s_1 s_2, \eta'n's'_1 s'_2} \equiv
   a_{\eta n s_1 s_2, \eta'n's'_1 s'_2}(E,E),
\label{A6}
\\ && \hspace{-1em}
   {\rm Re}\, A_{\eta n s_1 s_2} =
   \sum_{\eta' n' s'} a_{\eta n s_1 s_2, \eta'n's' s'},
   \label{A7}
\\ && \hspace{-1em}
   {\rm Im}\, A_{\eta n s_1 s_2} = 
   {1\over\pi}\sum_{\eta' n' s'} {\cal P} \int_{E_{n'}}^\infty
   {{\rm d}E'\over E'-E}
   a_{\eta n s_1 s_2, \eta'n's' s'}(E,E'),
\nonumber \\[-3ex] &&
\label{A8}
\end{eqnarray}
where $E_{n'} = \sqrt{1+2bn'}$,  
\begin{eqnarray}
&& \hspace{-1em}
   a_{\eta n s_1 s_2, \eta'n's' s''}(E,E') =
   {EE'\over|PP'|} \int_0^\infty {\rm d}u\,v^2(u,\xi) \times
\nonumber \\ &&
   ~~~~~~
   M^\ast_{s' s_1}(n',p'_z; n, p_z; u)
   M_{s'' s_2}(n',p'_z; n, p_z; u), 
\label{A9}
\end{eqnarray}
and $P\equiv P_{\eta n}(E)$ is defined in Sect.\,\ref{sect3.1}. 
According to Eqs.\,(\ref{3.26}) and
(\ref{3.2}), the matrix elements $M_{s's}$ contain the integrals
\begin{eqnarray}
&& \hspace{-1em}
  \int_{-\infty}^{\infty}
  {\cal H}_{n'}(y/ a_{\rm m} - q_x a_{\rm m}/ 2 )
  {\cal H}_n (y/a_{\rm m} + q_x a_{\rm m}/ 2)\, {\rm e}^{{\rm i}q_yy}
  {{\rm d}y\over a_{\rm m}} =
\nonumber \\[-1ex] && ~~~~
   \left[(n')!/n!\right]^{1/2}{\rm e}^{-u/2}u^{(n-n')/2}L_{n'}^{n-n'}(u)
   = I_{nn'}(u),
\label{A10}
\end{eqnarray}
where $I_{nn'}(u)$ is a Laguerre function (Sokolov \& Ternov 1968). 
Therefore $M_{s's}$ can be expressed in terms of the functions 
\begin{eqnarray}
   Q_2(\xi,n',n,m) & = &
   \int_0^\infty {I_{nn'}^2(u)\over (u+\xi)^m}\,
   {\rm e}^{-\zeta (u+\xi)}\,{\rm d}u ,
\label{A11}
\\
   Q_3(\xi,n',n,m) & \!= &
   \!\int_0^\infty {I_{nn'}(u)I_{n-1,n'-1}(u)\over (u+\xi)^m}
   \,{\rm e}^{-\zeta (u+\xi)}\,{\rm d}u ,
\label{A12}
\end{eqnarray}
and $Q_1=Q_2(\xi,n'-1,n-1,m)$. Here $m=1$ and 
$\zeta=2r_T^2/(3a_{\rm m}^2)$ for the scattering on phonons; $m=2$ and
$\zeta=0$ for the Coulomb scattering.
The functions $Q_1$, $Q_2$, and $Q_3$ have been introduced by
Yakovlev (1984) for the particular case of $\zeta=0$.
The properties of these functions are analysed in 
Appendix~B. 

Using the basis (\ref{3.2}),
we obtain
\begin{equation}
   a_{\eta n s_1 s_2, \eta'n's' s''}(E,E') =
   {S(s_1,s_2|s's'')\over 4(E+1)\,(E'+1)\,|PP'|},
\label{A13}
\end{equation}
where
\begin{eqnarray}
&& \hspace{-1em}
   S(1,1|1,1) = g^2Q_1+4b^2nn'Q_2+4bg\sqrt{nn'}\,Q_3,
\label{A14}
\\ && \hspace{-1em}
   S(-1,-1|-1,-1) = g^2Q_2+4b^2nn'Q_1+4bg\sqrt{nn'}Q_3,
\label{A15}
\\ && \hspace{-1em}
   S(1,1|-1,-1) =
\nonumber \\ &&
   ~~~~~~2b\,(n'P^2Q_1+nP'\,^2Q_2-2\sqrt{nn'}\,PP'Q_3),
\label{A16}
\\ && \hspace{-1em}
   S(-1,-1|1,1) =
\nonumber \\ &&
   ~~~~~~2b\,(n'P^2Q_2+nP'\,^2Q_1-2\sqrt{nn'}\,PP'Q_3),
\label{A17}
\\ && \hspace{-1em}
   S(-1,1|1,1) = S(1,-1|1,1) =  -g\sqrt{2bn}\,P'Q_1+
\nonumber\\ &&
   ~~~~~~2bn'\sqrt{2bn}PQ_2+\sqrt{2bn'}(gP-2bnP')Q_3,
\label{A18}
\\ && \hspace{-1em}
   S(1,1|-1,1) = S(1,1|1,-1) = -g\sqrt{2bn'}\,P Q_1+
\nonumber\\ &&
   ~~~~~~2bn\sqrt{2bn'}P'Q_2+\sqrt{2bn}(gP'-2bnP)Q_3,
\label{A19}
\\ && \hspace{-1em}
   S(-1,1|-1,-1) = S(1,-1|-1,-1) = g\sqrt{2bn}\,P'Q_2-
\nonumber\\ &&
   ~~~~~~2bn'\sqrt{2bn}PQ_1-\sqrt{2bn'}(gP-2bnP')Q_3,
\label{A20}
\\ && \hspace{-1em}
   S(-1,-1|-1,1) = S(-1,-1|1,-1) = g\sqrt{2bn'}\,P Q_2-
\nonumber\\ &&
   ~~~~~~2bn\sqrt{2bn'}P'Q_1-\sqrt{2bn}(gP'-2bnP)Q_3,
\label{A21}
\\ && \hspace{-1em}
   S(-1,1|-1,1)  = S(1,-1|1,-1) =
\nonumber \\ &&
   ~~~~~~2bg\sqrt{nn'}\,(Q_1+Q_2)+(4b^2nn'+g^2)\,Q_3,
\label{A22}
\\ && \hspace{-1em}
   S(-1,1|1,-1)  = S(1,-1|-1,1) =
\nonumber \\ &&
   ~~~~~~2b\sqrt{nn'}\,PP'(Q_1+Q_2)-2b\,(n'P^2+nP'\,^2)Q_3,
\label{A23}
\end{eqnarray}
and
%
$g=(E+1)\,(E'+1)+PP'$.
%

The first two pairs of coefficients (Eqs.\,(\ref{A14}), (\ref{A15}),   
and (\ref{A16}), (\ref{A17})),
are related to diagonal density matrix elements
for electron transitions without and with spin flip, respectively.
At $E=E'$ they reproduce equations of Yakovlev (1984). 
The remaining equations (\ref{A18}) -- (\ref{A23}) 
present the coefficients at off-diagonal 
elements of the correction $\varphi$ to the density matrix 
in the system (\ref{3.29}). 
\addtocounter{section}{1}
\section*{Appendix B: integrals $Q_i(\xi,n',n,m)$}       
\setcounter{equation}{0}
For an efficient computation of the coefficients (\ref{A6}) --
(\ref{A8}) in Eqs.~(\ref{3.29}),
let us consider the properties of the functions $Q_i$
(Eqs.\,(\ref{A11}), (\ref{A12})). Since
$Q_i(\xi,n',n,m) = Q_i(\xi,n,n',m)$, we assume $n'-n \geq 0$
without loss of generality.

Using the polynomial representation for the Laguerre polynomials,
the integrals $Q_i$ can be expressed as 
\begin{eqnarray}
   Q_2(\xi,n',n,m) &=& \sum_{j=0}^{2n} (-1)^j \sum_k c_{n'nk}
   c_{n'n,j-k}
   \times
\nonumber\\ &&
   (n'-n+j)!\,Q_2(\xi,n'-n+j,0,m),
\label{B01}
\\
   Q_3(\xi,n',n,m) &=& \sum_{j=0}^{2n-1} (-1)^j \sum_k c_{n'nk}
   c_{n'n,j-k}{n-k\over\sqrt{n'n}}
   \times
\nonumber\\ &&
   (n'-n+j)!\,
    Q_2(\xi,n'-n+j,0,m),
\label{B02}
\\
   Q_2(\xi,n',0,1) &=& 
   (1+\zeta)^{-n'}\,{\rm e}^\xi\,E_{n'+1}(\xi+\zeta\xi), 
\label{B03}
\\
   Q_2(\xi,n',0,2) &=& {\rm e}^\xi 
   \,\left(E_{n'}(\xi) - E_{n'+1}(\xi)\right), 
\label{B03a}
\end{eqnarray}
where $E_i(\xi)$ is an integral exponent which is 
easily calculated (Abramowitz \& Stegun 1972), 
\begin{equation}
   c_{n'nk} = {\sqrt{n'!n!}\over k!(n-k)!(n'-n+k)!}
\label{B04}
\end{equation}
are the coefficients of the polynomial expansion of 
a Laguerre function $I_{n'n}(u)$, 
and summation index $k$ runs from $\max(0,j-n)$ to
$\min(n,j)$. 
However this method fails for $n\ga 15$ due to 
the exponentially increasing round-off errors 
(Hernquist 1984, Schaaf 1988).
Direct integration (Hernquist 1984) is not very efficient 
because of rapid oscillations of the Laguerre functions at large 
$n$. The direct integration is especially undesirable for the 
density matrix computation, since an outer integration over $E'$ 
is required in Eq.\,(\ref{A8}). 

To avoid the above difficulties, we propose to use
the transformation
\begin{equation}
   (u+\xi)^{-m} = \int_0^\infty
   {x^{m-1}\over(m-1)!}\,{\rm e}^{-(u+\xi)\,x}{\rm d}x
\label{B1}
\end{equation}
in Eqs.\,(\ref{A11}), (\ref{A12}), 
and change then the integration order. Let us use the
equalities (Gradshtein \& Ryzhik 1965, Abramowitz and Stegun 1972)
\begin{eqnarray}
   && \hspace{-1em}
   \int_0^\infty u^{n'-n}{\rm e}^{-u\,(1+x)}
   L_n^{n'-n}(u)\,L_{n-j}^{n'-n}(u)\,{\rm d}u =
\nonumber\\
   [-.5ex] &&~~~~~~
   {(n'+n-j)!\,x^{2n-j} \over (n'-n)!\,(n-j)!\,n!\,(1+x)^{n'+n-j+1}}
   \times
\nonumber\\
   &&~~~~~~
   F\left(-n+j,-n; -n'-n+j; 1- {1/x^2}\right) =
\label{B2}\\
    &&
   {x^j\over(1+x)^{n+n'-j+1}} \sum_{k=0}^{n-j}
   {(n')!\,(n'-j)! \over (n-j-k)!\,(n'-j-k)!(j+k)!}\,{x^{2k}\over k!},
\nonumber
\end{eqnarray}
where $F$ is a Gauss hypergeometric function.
Then we obtain
\begin{eqnarray}
   && \hspace{-1em}
   Q_{2+j}(\xi,n',n,m) = \sum_{k=0}^{n-j}
   {\left[n!\,(n')!\,(n-j)!\,(n'-j)!\right]^{1/2} \over
   k!\,(n-j-k)!\,(n'-j-k)!} \times
\nonumber\\
   && {1\over(m-1)!\,(j+k)!}\,
   \int_\zeta^\infty {x^{2k+j+m-1}\over(1+x)^{n'+n-j+1}}\,
   {\rm e}^{-\xi x}\,{\rm d}x.
\label{B3}
\end{eqnarray}
The main advantage of the representation 
(\ref{B3}) is that all terms are positive monotonous
functions of $\xi$.

Using Eq.\,(\ref{B3}), we can easily derive various asymptotes.
In particular, if $\zeta=0$ and 
$\xi\to\infty$, then from the asymptotic 
properties of the integral in Eq.\,(\ref{B3}) 
we obtain 
\begin{eqnarray}
   && \hspace{-1em}
   Q_{2+j}(\xi,n',n,m) \simeq
   \left[{n!\,(n')!\over(n-j)!\,(n'-j)!}\right]^{1/2}\,
   {(j+m-1)!\over j!\,(m-1)!}\times
\nonumber\\
   &&~~~~~~
   (1/ \xi)^{m+j}\,\left[1-(n+n'-j+1)\,(j+m)/\xi+\ldots\right].
\end{eqnarray}

The functions $Q_i$ are finite for $n'\neq n$, and they diverge
at $\xi\to 0$ for $n'=n$:
\begin{eqnarray}
&& \hspace{-1em}
         Q_2(\xi,n,n,1) = -\ln\,\xi-\gamma-\sum_{m=1}^{2n}{1\over m}+
    \nonumber\\ && 
         {(n!)^2\over (2n)!} \sum_{m=1}^n 
         {(2n-2m)!\,(2m-1)! \over [(n-m)!\,m!\,]^2} + O(\xi ),
\label{B4} \\
&& \hspace{-1em}
         Q_3(\xi,n,n,1) = -\ln\,\xi-\gamma-\sum_{m=1}^{2n-1}{1\over m}+
    \nonumber\\ && 
         {n!\,(n-1)!\over (2n-1)!} \sum_{m=1}^{n-1} 
         {(2n-2m-1)!\,(2m-1)! \over (n-m)!\,(n-m-1)!\,(m!)^2} + O(\xi ),
\label{B4a} \\
&& \hspace{-1em}
   Q_2(\xi,n,n,2) = \xi^{-1} + 
   (2n+1)\,\left[\ln\,\xi+\sum_{m=2}^{2n+1}{1\over m} + \gamma \right] + 
    \nonumber\\ && 
         {(n!)^2\over (2n)!} \sum_{m=1}^{n} 
         {(2n-2m+1)!\,(2m-2)! \over [(n-m)!\,m!\,]^2} + O(\xi ),
\label{B5} \\
&& \hspace{-1em}
   Q_3(\xi,n,n,2) = \xi^{-1} + 
   2n\,\left[\ln\,\xi+\sum_{m=2}^{2n}{1\over m} + \gamma \right] + 
    \nonumber\\ && 
         {n!\,(n-1)!\over (2n-1)!} \sum_{m=1}^{n-1} 
         {(2n-2m)!\,(2m-2)! \over (n-m)!\,(n-m-1)!\,(m!)^2} + O(\xi ). 
\label{B6}
\end{eqnarray}
However, collecting the terms with $n'=n$ in Eq.\,(\ref{3.29}), 
we see that only the finite terms of 
Eqs.\,(\ref{B4})--(\ref{B6}) do contribute to the coefficients 
of the algebraic system, while 
contributions from the divergent terms 
($\xi^{-1}$ and $\ln{\xi}$) 
are mutually compensated in the real parts of the coefficients 
(Eqs.\,(\ref{A6}), (\ref{A7})), as well as in the integrals (\ref{A8}) 
for the imaginary parts. Therefore the coefficients of the 
density matrix equations are essentially finite, except for the 
singularities at the Landau thresholds due to the factor $|PP'|$ in the 
denominators of Eqs.\,(\ref{A9}), (\ref{A13}). These remaining 
singularities stem from divergencies of the electron 
density of states at the Landau thresholds, and they are responsible for 
the quantum oscillations of the relaxation time 
discussed in Sect. \ref{sect3.5}. 

Finally, consider the case $n\gg 1$. Let us make 
use of the semiclassical approximation 
of a Laguerre function averaged over oscillations 
(Kaminker \& Yakovlev 1981): 
\begin{equation}
   \overline{I^2_{nn'}(u)} \approx
   \left[\pi\sqrt{(u-u_1)\,(u-u_2)}\right]^{-1},
\label{B8}
\end{equation}
where $u_{1,2} = (\sqrt{n'}\mp\sqrt{n})^2$. 
Using again the transformation (\ref{B1}), we express the integrals
$Q_i$ as
\begin{eqnarray}
&& \hspace{-1em}
   Q_2(\xi,n',n,1) = 
   {1\over\pi}\int_0^\pi 
   {\exp\left(-2\zeta\sqrt{n'n}\,\cos\vartheta\right) \over 
   n'+n+\xi+2\sqrt{n'n}\,\cos\vartheta}\,{\rm d}\vartheta
\label{B8a}\\&&
   ={1\over(n'+n+\xi)}\int_{\zeta(n'+n+\xi)}^\infty\!\!
   {\rm d}x\,{\rm e}^{-x}
   I_0\left({2x\sqrt{n'n}\over n'+n+\xi}\right),
\label{B8aa}\\
   && \hspace{-1em}
   4\sqrt{n'n}\,Q_3(\xi,n',n,1) =
   (n+n')\,\left[Q_2(\xi,n',n,1)+\right.
\nonumber\\
   && \left.
   Q_2(\xi,n'-1,n-1,1)\right] +
   \xi\,\left[Q_2(\xi,n',n-1,1)+\right.
\nonumber \\
   && \left.
   Q_2(\xi,n'-1,n,1)\right]  -
   \exp\left(-\zeta(n'+n+\xi-1)\right)\times
\nonumber\\
   && \left[I_0\left(2\zeta\sqrt{n'(n-1)}\right)+
   I_0\left(2\zeta\sqrt{(n'-1)n}\right)\right],
\label{B8b}\\
   && \hspace{-1em}
   Q_i(\xi,n',n,2) = -(\partial/\partial\xi)\,Q_i(\xi,n',n,1),
\label{B8c}
\end{eqnarray}
where $I_0$ is a modified Bessel function.
In deriving Eq.\,(\ref{B8b}), we have used
the relationship (Kaminker \& Yakovlev 1981)
\begin{eqnarray}
   && \hspace{-1em}
   4\sqrt{nn'}\,I_{nn'}(u)\,I_{n-1,n'-1}(u) =
   (n+n')\,\left(I^2_{nn'}(u)+
   \right.
\nonumber\\
   && ~~~~~~ \left. I^2_{n-1,n'-1}(u)\right) -
   u\,\left(
   I^2_{n-1,n'}(u)+I^2_{n,n'-1}(u)
   \right).
\label{B13}
\end{eqnarray}
The integral (\ref{B8aa}) can be easily calculated using 
polynomial approximations for $I_0(x)$ (Abramowitz \& Stegun 1972). 
At $2\sqrt{n'n}/(n'+n+\xi) < 0.95$, however, 
direct integration of Eq.\,(\ref{B8a}) is more efficient, 
since only a few integration mesh points are needed then. 
If $\zeta=0$, the semiclassical approximation 
allows us to express $Q_i$ in elementary functions: 
\begin{eqnarray}
   && \hspace{-1em}
   Q_2(\xi,n',n,1) = \left[(u_1+\xi)\,(u_2+\xi)\right]^{-1/2},
\label{B9} \\
   && \hspace{-1em}
   Q_2(\xi,n',n,2) = (\xi+n+n')\,
   \left[(u_1+\xi)\,(u_2+\xi)\right]^{-3/2},
\label{B10}
         \\ && \hspace{-1em}
         4\sqrt{nn'}\,Q_3(\xi,n',n,1) = 
    \nonumber\\ && 
         (n+n')\,(Q_2(\xi,n',n,1)+Q_2(\xi,n'-1,n-1,1))+
    \nonumber\\ && 
         \xi\,(Q_2(\xi,n',n-1,1)+Q_2(\xi,n'-1,n,1))-2, 
\label{B11}
         \\ && \hspace{-1em}
         4\sqrt{nn'}\,Q_3(\xi,n',n,2) = -((n+n')\xi+(n'-n)^2)\,\times 
    \nonumber \\ && 
         \left(Q_2^3(\xi,n'-1,n,1)+Q_2^3(\xi,n',n-1,1)\right) + 
    \nonumber\\ && 
         (n+n')\,(Q_2(\xi,n',n,2)+Q_2(\xi,n'-1,n-1,2)). 
\label{B12}
\end{eqnarray}
We have checked numerically that errors 
in calculating $\Phi(\epsilon)$ do not exceed 3\%, 
if we substitute 
the semiclassical formulae (\ref{B8a}) -- (\ref{B8b}) and 
(\ref{B9}) -- (\ref{B12}) into Eqs.\,(\ref{A14}) -- (\ref{A23}) 
for $n>10$. The semiclassical approximation makes the 
computation much faster. 
\addtocounter{section}{1}
\section*{Appendix C: a fitting formula to electron density} 
\setcounter{equation}{0}
Equation (\ref{2.0}) can be represented as 
\begin{equation}
   n_{\rm e} = {m_{\rm e}\omega_B k_{\rm B}T\over 2(\pi\hbar)^2 c} 
   \sum_{n,s} F\left({\mu-\epsilon_n \over k_{\rm B}T},
   {\epsilon_n \over k_{\rm B}T}\right), 
\label{C1}
\end{equation}
where $\epsilon_n\equiv\epsilon_n(0)$ is defined 
by Eq.\,(\ref{2.0a}) with $p_z=0$, and 
\begin{equation}
   F(x,y) = 
   \int_0^\infty {{\rm e}^{t-x} \over ({\rm e}^{t-x}+1)^2}\,
   \sqrt{t(t+2y)}\,{\rm d}t. 
\label{C2}
\end{equation}
Evidently, 
\begin{equation}
   {\partial n_e\over\partial\mu} = 
   {m_{\rm e}\omega_B\over 2(\pi\hbar)^2 c}
   \sum_{n,s} F'\left({\mu-\epsilon_n \over k_{\rm B}T},
   {\epsilon_n \over k_{\rm B}T}\right), 
\label{C2a}
\end{equation}
where $F'(x,y)\equiv \partial F(x,y)/\partial x$. 

In the non-relativistic limit we have $y\gg 1$ and $x\ll y$ 
(or $x<0$). Then $F(x,y)$ 
reduces to $\sqrt{y/2}\,{\cal F}_{-1/2}(x)$, where 
${\cal F}_{-1/2}(x)$ is a Fermi integral, for which some useful 
approximations have been presented by Antia (1993). In the 
opposite case of $y\ll 1$ we have 
$F(x,y)=\ln\left(1+{\rm e}^x\right)$. 

In the general case of arbitrary $y$, the following approximation 
is proposed: 
\begin{equation}
   F(x,y) = \ln\left(1+{\rm e}^x\right)\,
   {1+y+\xi+c(y)\,a(\xi)\,\sqrt{\xi+2y} \over 
   1+\xi+c(y)\,b(\xi)}, 
\label{C3}
\end{equation}
where 
$\xi = \ln(1+\exp[x-x_0(y)])$, \\[.3ex] 
$x_0(y) = \left(1+c_1 y^{c_2}\right)^{-1}$, \\[.3ex]
$c(y) = c_3 y^{c_4}$, \\[.3ex]
$a(\xi) = \sqrt{\pi}\,/2 + (a_1 + a_2\xi^2)\,\sqrt{\xi}$, \\[.3ex]
$b(\xi) = 1 + a_3\sqrt{\xi} + a_4\xi + a_2\xi^3$, \\[.3ex]
with the numerical parameters 
$c_1=0.623$, $c_2=1.6031$, $c_3=0.9422$, $c_4=1.7262$, 
$a_1 = 0.103$, $a_2 = 0.043$, $a_3 = 0.0802$, and $a_4 = 0.2944$. 
At $y\gg1$, the right-hand side of Eq.\,(\ref{C3}) 
simplifies to 
$\xi\,\sqrt{\xi+2y}\,a(\xi)/b(\xi)$, with $\xi=\ln(1+{\rm e}^x)$, 
and depends essentially only on the four constants $a_i$. 
An error of the approximation (\ref{C3}) reaches maximum of 
0.57\% at $x=2.9,~ y=5.9$. We have checked additionally, that 
the $x$-derivative of this approximation fits the function $F'(x,y)$, 
which enters Eq.\,(\ref{C2a}), 
with a maximum relative error of 2\% . 

\newpage
\onecolumn
\section*{Figure captions}
\begin{figure*}[h]
\caption[ ]{
Function $\Phi(\epsilon)$ given by Eq.\,(\ref{3.16}).
Landau variable $\nu$ is defined by Eq.\,(\ref{3.17}). \\
Top panels: $\Phi(\epsilon)$ in the density matrix approach. \\
Bottom panels: Relative errors 
$\Delta\Phi/\Phi = (\Phi_{\rm appr}/\Phi-1)$  
of the distribution function approximations $\Phi_{\rm appr}$ 
(dotted line: basis (\ref{3.2}); dot-and-dashed line: basis (\ref{3.4}),
(\ref{3.5})).  Dashes show non-magnetic approximation, according to 
Eqs.\,(\ref{4.10b}), (\ref{4.10c}), and (\ref{4.1}).
}
\label{fig1}
\end{figure*}
\begin{figure*}[h]
\caption[ ]{
Ratios of the effective relaxation times 
$\tau_{\rm DW}(\epsilon)$ to $\tau_{\rm non-DW}(\epsilon)$ 
calculated, respectively, with 
and without the Debye -- Waller factor. \\
The temperature $T = T_7\times 10^7$~K, with $T_7$ = 0.2, 0.5, 1, 2, 
5, and 10. 
Solid lines -- numerical results (Eqs.\,(\ref{4.1}), (\ref{3.16})), 
dashed lines -- non-magnetic case (Eqs.\,(\ref{4.10b}), 
(\ref{4.10c})). The curves are plotted starting from the melting 
points. } 
\label{fig2}
\end{figure*}

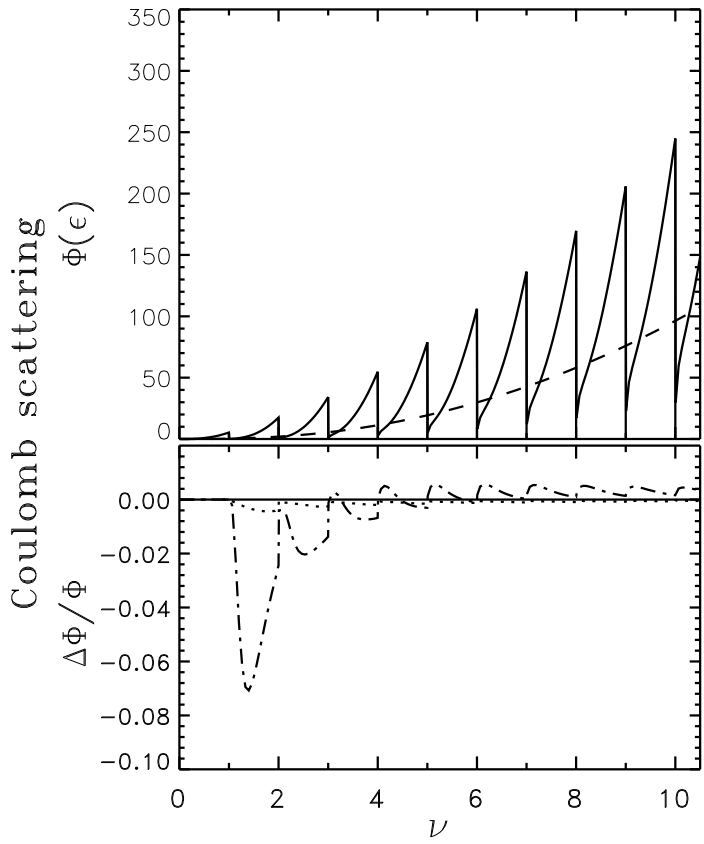
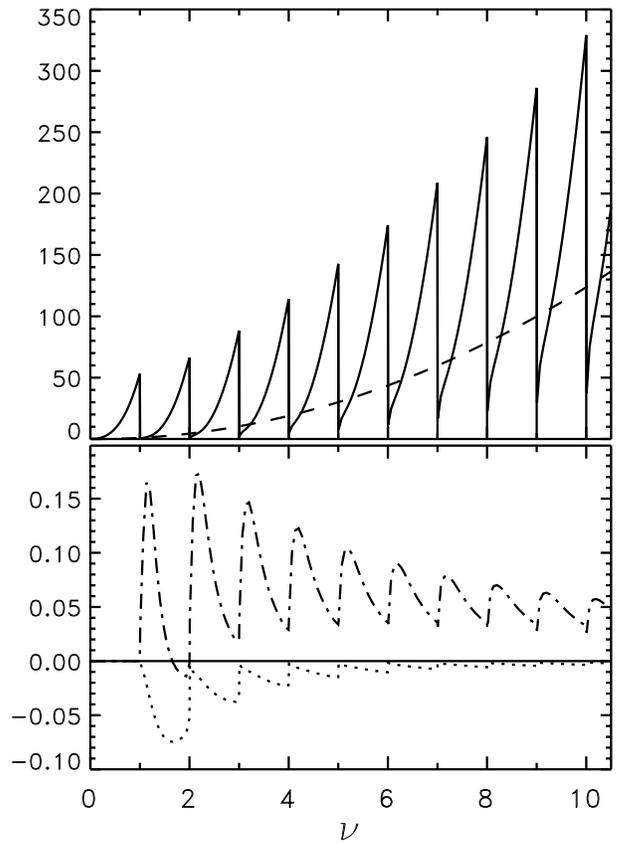
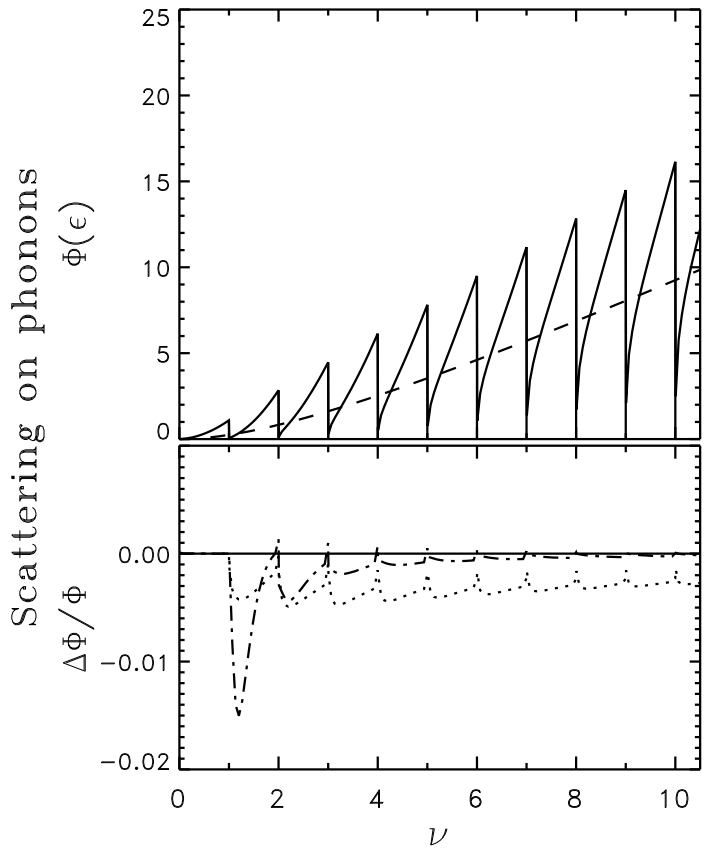
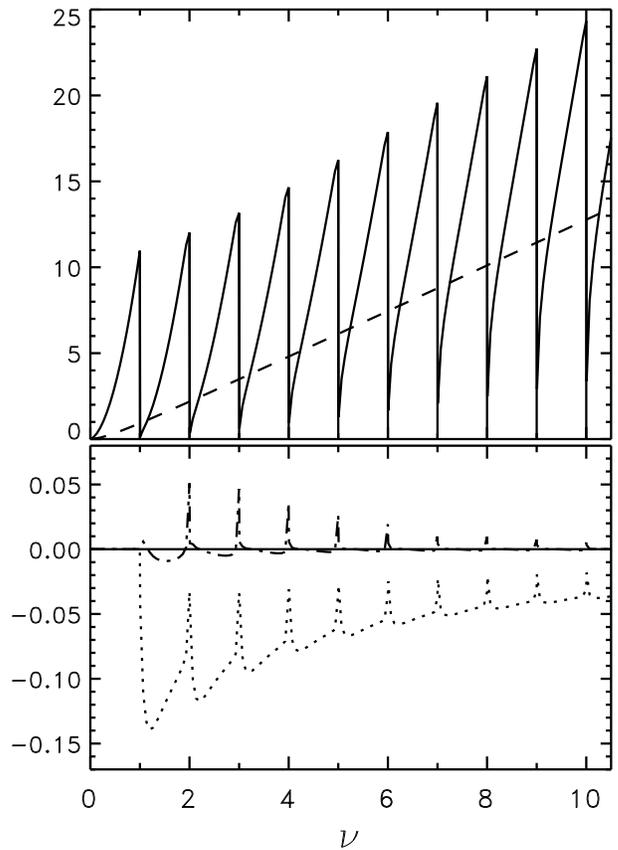
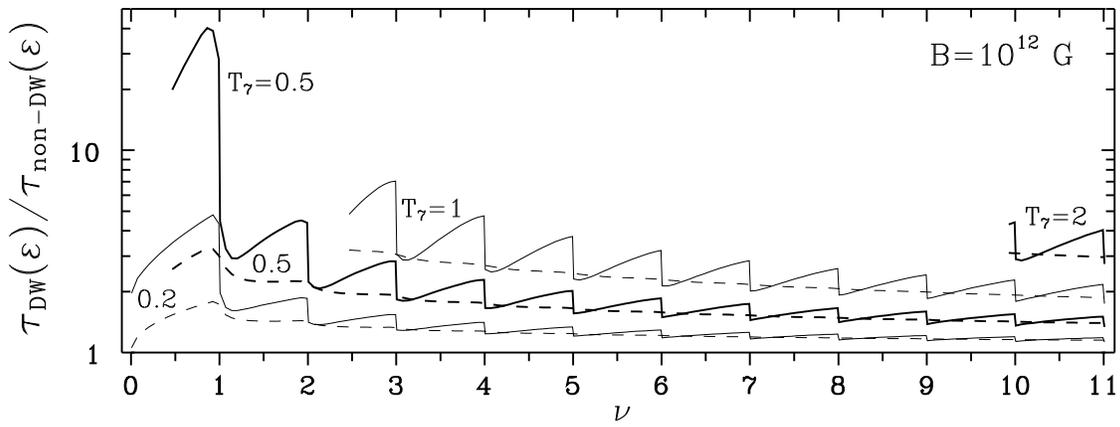
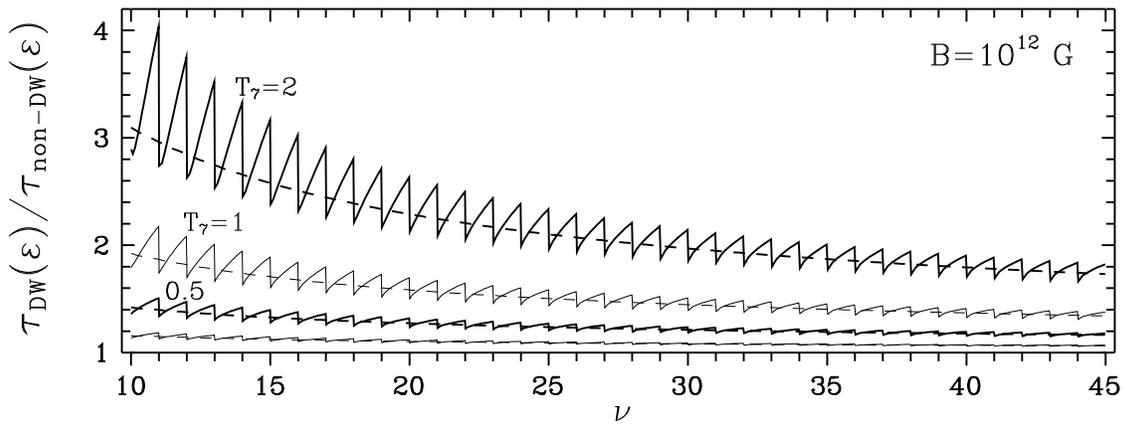
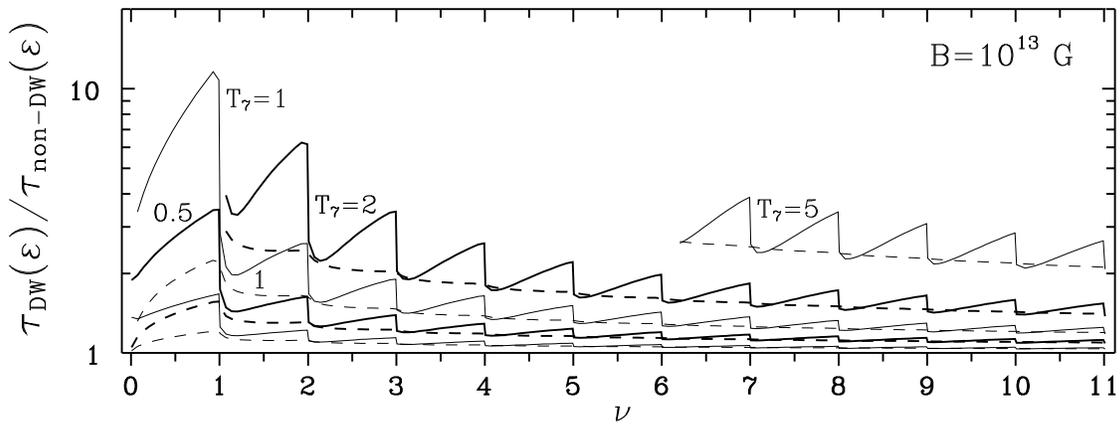
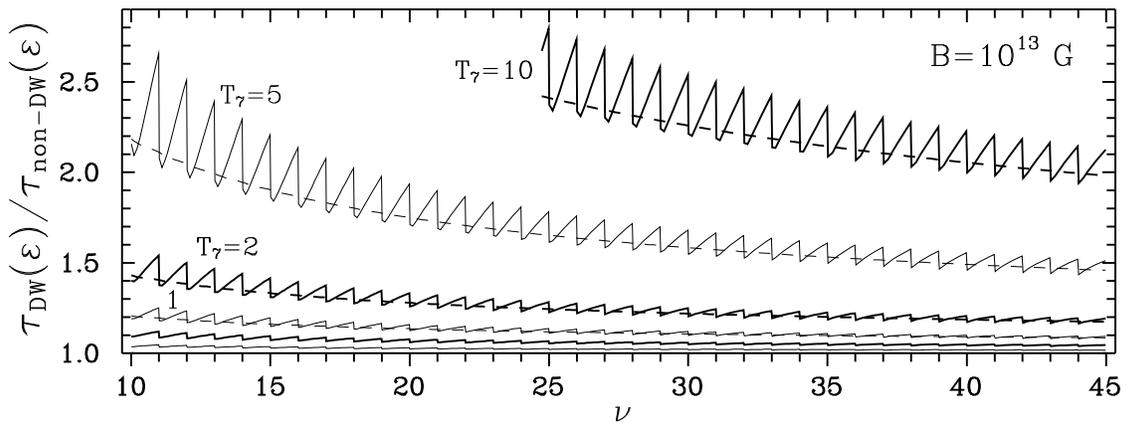
\end{document}